\documentclass[12pt,journal,final,onecolumn]{IEEEtran}

\usepackage{graphicx}
\usepackage{color}
\usepackage[cmex10]{amsmath}
\usepackage{amssymb}
\usepackage{latexsym}
\usepackage{amsthm}
\usepackage{amsfonts}
\usepackage{bm}
\usepackage{cite}
\usepackage[tight,footnotesize]{subfigure}
\usepackage{url}
\usepackage{booktabs}

\graphicspath{{figs/}}

\newcommand{\mc}[1]{\mathcal{#1}}

\newcommand{\mcV}{\mc{V}}
\newcommand{\mcA}{\mc{A}}
\newcommand{\mcP}{\mc{P}}

\newcommand{\defeq}{\mathrel{\triangleq}}
\newcommand{\Pp}{\mathbb{P}}
\newcommand{\E}{\mathbb{E}}

\newcommand{\Z}{\mathbb{Z}}

\newcommand{\R}{\mathbb{R}}

\newcommand{\ceil}[1]{\lceil{#1}\rceil}

\newcommand{\abs}[1]{\lvert{#1}\rvert}
\newcommand{\card}[1]{\abs{#1}}

\newcommand{\iid}{i.\@i.\@d.\ }

\DeclareMathOperator{\out}{out}
\DeclareMathOperator{\tail}{tail}
\DeclareMathOperator{\id}{id}

\newtheorem{lemma}{Lemma}

\newtheorem{theorem}[lemma]{Theorem}

\newtheorem{innercustomtheorem}{Theorem}
{\renewcommand\theinnercustomtheorem{#1}\innercustomtheorem}%
{\endinnercustomtheorem}

\theoremstyle{definition}

\newtheorem{egdummy}{Example}
\newenvironment{example}[1][]%
{%
    \begin{egdummy}[#1]%
    \upshape%
}%
{%
    \qed%
    \end{egdummy}%
}

\newtheoremstyle{myremark}%
{\topsep}{\topsep}{\normalfont}{\parindent}{\itshape}{:}{ }{}

\theoremstyle{myremark}
\newtheorem{remark}{Remark}

\newcommand\shortintertext[1]{%
\ifvmode\else\\\@empty\fi
\noalign{%
\penalty0%
\vbox{\mathstrut}%
\penalty10000%
\vskip-\baselineskip
\penalty10000%
\vbox to 0pt{%
\normalbaselines
\ifdim\linewidth=\columnwidth
\else
\parshape\@ne
\@totalleftmargin\linewidth
\fi
\vss
\noindent#1\par}%
\penalty10000%
\vskip-\baselineskip}%
\penalty10000}

\begin{document}

\title{On the Problem of Optimal Path Encoding \\ for Software-Defined Networks}

\author{Adiseshu Hari, Urs Niesen, Gordon Wilfong%
    \thanks{Author names in alphabetical order.}%
    \thanks{Adiseshu Hari and Gordon Wilfong are with Bell Labs, Nokia. Emails:
    \{adiseshu.hari, gordon.wilfong\}@nokia.com. Urs Niesen was
    with Bell Labs; he is now with Qualcomm's NJ Research
    Center. Email: urs.niesen@ieee.org}%
    \thanks{This work was presented in part at the IEEE International
    Symposium on Information Theory, June 2015.}%
}

\maketitle

\begin{abstract} 
    Packet networks need to maintain state in the form of forwarding
    tables at each switch. The cost of this state increases as networks
    support ever more sophisticated per-flow routing, traffic
    engineering, and service chaining.  Per-flow or per-path state at
    the switches can be eliminated by encoding each packet's desired
    path in its header.  A key component of such a method is an
    efficient encoding of paths through the network.  We introduce a
    mathematical formulation of this optimal path-encoding problem. We
    prove that the problem is APX-hard, by showing that approximating it
    to within a factor less than $8/7$ is NP-hard. Thus, at best we can
    hope for a constant-factor approximation algorithm. We then present
    such an algorithm, approximating the optimal path-encoding problem
    to within a factor $2$.  Finally, we provide empirical results
    illustrating the effectiveness of the proposed algorithm.
\end{abstract}

\section{Introduction}
\label{sec:intro}

New networking technologies such as network
virtualization~\cite{network-virtualization, Geni}, policy-based
routing~\cite{pathlet, qazi2013simple}, per-flow
routing~\cite{sen2013scalable}, and service
chaining~\cite{service-chaining} are leading to an explosion of state
maintained at switches in the network core. Current efforts
to control this state rely on restricting the per-flow state to the
network edges and using tunnels in the network core, which only
maintains the traditional per-destination forwarding state. While
solving the state problem, this approach results in suboptimal routing,
because multiple flows, each with potentially different delay, jitter,
and bandwidth requirements, are aggregated into a single tunnel. 

One way of extending per-flow state from the network edge to the core is
for the header of each packet to contain an encoding of the packet's
required path. One such approach, termed source routing, encodes the
path as a sequence of identifiers (such as the IP addresses) of the
intermediate hops along the path~\cite{sunshine77}. This approach
results in large packet headers and still requires each switch along the
path to perform a table lookup to translate the identifier to an
interface (i.e., an output port) from a table that grows as the size of
the network grows.

\begin{figure}[htbp]
    \centering 
    \includegraphics{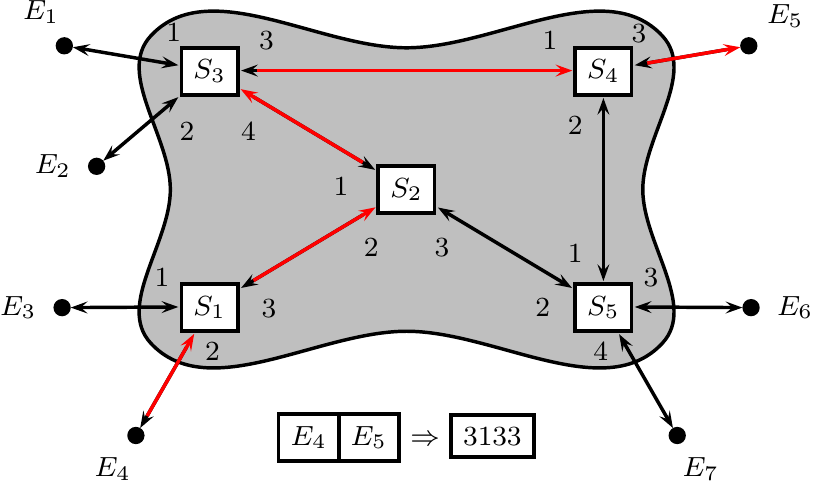} 
    \caption{Packet traversal using the path-encoding architecture.}
    \label{fig:overview}
\end{figure}

To reduce the size of state, i.e., the size of the lookup table at
each core switch, one can instead encode the path as a sequence of
interface identifiers~\cite{path-addressing, soliman12}.  For instance,
if a switch has $k$ interfaces, then each of them could be assigned a
distinct label of length $\ceil{\log k}$ and a path encoding is a
sequence of such labels. We refer to this approach as path switching.

For example, consider the scenario depicted in Fig.~\ref{fig:overview},
in which an incoming packet to the network from endpoint $E_4$ to
endpoint $E_5$ has its destination encoded at the ingress switch $S_1$
as $3133$.  This encoded path represents a sequence of labels, each
uniquely identifying a switch interface on the path between the source
and the destination. In our example, the first switch $S_1$ forwards the
packet through its interface $3$ to switch $S_2$; $S_2$ forwards the
packet through its interface $1$ to switch $S_3$; $S_3$ forwards the
packet through its interface $3$ to switch $S_4$; and so on. 

Encoding a packet's path in this manner eliminates expensive lookup
tables at the switches in the network core, since each switch can
identify the next hop from the interface label within the encoded path.
Thus, path encoding enables arbitrary routing without the need to
maintain per-path or per-flow lookup tables in the core switches. In
particular, paths through the network can be arbitrarily complicated,
making this approach ideally suited for service chaining and
sophisticated traffic engineering with per-flow granularity.  With such
an encoding, the size of a switch's lookup table to translate a label to
an interface depends only on the number of interfaces at
that switch and therefore remains constant despite network growth.

In this paper, we propose a path encoding that, rather than constraining
the interface labels at a given switch to have the same length, allows
the interface labels at the same switch to have variable lengths. The
flexibility of variable-length interface encoding advocated in this
paper has the advantage of resulting in shorter encoded paths. In
particular, our contribution is a method for path encoding that
minimizes the maximum length of any encoded path in the network.
Minimizing this performance measure is appropriate when each encoded
path is to be placed inside existing source/destination address header
fields. As with any reasonable encoding using a sequence of interface
labels, the proposed encoding allows the state kept at each switch to
remain of constant size independent of the growth of the network in
terms of topology as well as in terms of the number of distinct flows.
Finally, the proposed method allows each switch along a path to easily
and unambiguously determine the correct outgoing interface.

We begin by introducing a mathematical formulation of the problem of
computing interface labels that minimize the longest encoded path for a
given set of paths. We call this the \emph{optimal path-encoding
problem}. We prove that this problem is APX-hard, by showing that it is
NP-hard to approximate to within any factor better than $8/7$ of
optimal. We next describe a $2$-approximation algorithm for the optimal
path-encoding problem. Finally, we apply the proposed algorithm to
several real-world networks (the AT\&T MPLS backbone
network~\cite{att2008} and 11 networks from the RocketFuel topology
set~\cite{rocketfuel}). The proposed variable-length encoding results in
a reduction of up to $30\%$ in the maximum encoded path length.

The remainder of this paper is organized as follows.
Section~\ref{sec:formulation} introduces the optimal path-encoding
problem. Algorithmic solutions to this problem are presented in
Sections~\ref{sec:main} and~\ref{sec:proofs}. Finally,
Section~\ref{sec:conclusion} contains concluding remarks.

\section{Problem Formulation}
\label{sec:formulation}

We start with a more detailed description of the network architecture in
Section~\ref{sec:architecture}. A key component of this architecture is
the optimal encoding of paths described in Section~\ref{sec:problem}. We
introduce a mathematical formulation of this problem in
Section~\ref{sec:math}.

\subsection {Network Architecture}
\label{sec:architecture}

Consider the conceptual model of a network shown in
Fig.~\ref{fig:overview}. Endpoints are connected to the network via an
edge switch. The network is assumed to be software-defined networking
(SDN) enabled, i.e., there is a centralized SDN controller that
configures each switch in the network. The SDN controller installs an
interface-label table in each switch, assigning to each
(outgoing) switch interface a unique binary string called the interface
label. In addition, the SDN controller installs a flow-table in each
edge switch, assigning to each incoming flow an encoded path. Each such
encoded path is a binary string, consisting of the concatenation of the
interface labels in the path.

Edge switches use the flow-table entries to modify the packet headers of
incoming and outgoing flows. How this mapping between flows and encoded
paths is performed depends on the nature of the packet endpoints.  If
the network ingress and egress interfaces in the path uniquely identify
the packet endpoints, then the encoded path replaces the source and
destination fields. Otherwise, the encoded path is placed in a tunnel
header, leaving the existing packet header unchanged. 

To facilitate the forwarding operations inside the network, the path
label also contains a pointer field indicating the current position in
the encoded path. In order to forward a packet, the switch reads its path
label starting from the position of the pointer. It then searches its
interface-label table for the unique interface that could result in this
encoded path (starting from the current position). It increments the
pointer by the length of the label, and forwards the packet over the
corresponding interface.

From the above description, we see that a key component of this
architecture is the assignment by the SDN controller of binary labels to
switch interfaces. This assignment needs to be such that switches are
able to make correct forwarding decisions. At the same time, the
resulting encoded paths cannot be too large so that they can fit inside
existing packet source/destination address fields, thereby leaving
packet formats invariant. We next describe this path-encoding problem in
detail.

\subsection{The Optimal Path-Encoding Problem}
\label{sec:problem}

As mentioned above, the goal of path encoding is the assignment of
labels to switch interfaces such that the following two objectives are
satisfied. First, to ensure that packets can be properly routed, the
encoded paths need to be uniquely decodable. Second, because the
encoding is appended to every packet sent through the network, usually
in a header of fixed size, the longest encoded path needs to be small.  

\begin{figure}[htbp]
    \centerline{%
    \hfill%
    \includegraphics{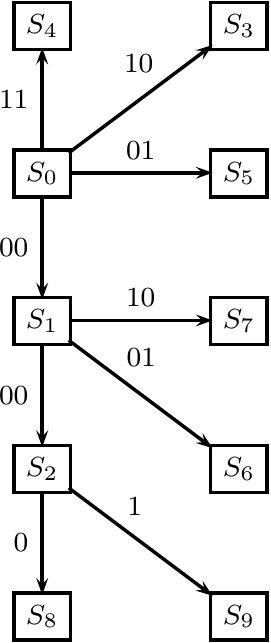}%
    \hfill%
    \includegraphics{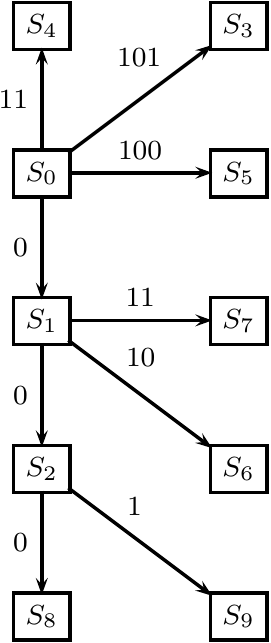}%
    \hfill%
    }
    \caption{Path encoding with fixed-length labels (left) and
    variable-length labels (right). The vertices represent
    switches and the arcs interfaces. The figure shows next to each 
    arc the label of the corresponding switch interface. 
    The set of paths are those from the switch $S_0$ to all the edge
    switches $S_3, S_4, \dots, S_9$.}
    \label{fig:encoding1}
\end{figure}

One way to solve the path-encoding problem is to assign fixed-length
labels to each switch interface. More precisely, for a switch with $k$
interfaces, we can assign a binary label of size $\ceil{\log k}$ bits to
each of its interfaces. As an example of this fixed-length labeling,
consider the network and label assignment depicted on the left side of
Fig.~\ref{fig:encoding1}. In this example, we consider all possible
paths from switch $S_0$ to any of the edge switches $S_3, S_4, \dots,
S_9$. The longest encoded path is $(S_0, S_1, S_2, S_8)$ with encoding
$00000$ of length $5$. Observe that, for ease of presentation, we do not
explicitly distinguish between switches and endpoints in this example
and in the following.

In this work, we instead advocate the use of variable-length interface
labeling, in which labels for interfaces of the same switch may have
different lengths. As an example of this variable-length labeling, consider
the same network as before, but with the label assignment depicted on
the right side of Fig.~\ref{fig:encoding1}. Assuming the same set of
paths as before, one longest encoded path is again $(S_0, S_1, S_2,
S_8)$, but this time, its encoding $000$ has only length $3$. Thus, we
see that in this example the use of variable-length labels has 
reduced the longest encoded path from $5$ to $3$ bits.

\begin{figure}[htbp]
    \centerline{%
    \hfill%
    \includegraphics{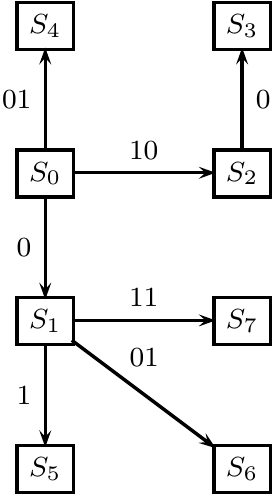}%
    \hfill%
    \includegraphics{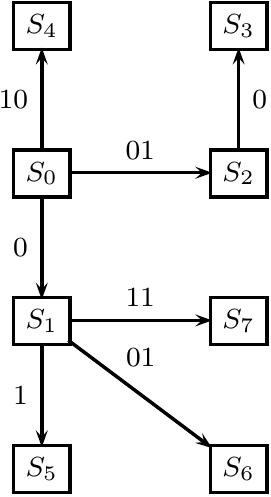}%
    \hfill%
    }
    \caption{Incorrect path encoding with variable-length labels leading
        to encoded paths that are not unique (left) or that are not
        locally decodable (right). The set of paths are those 
        from the switch $S_0$ to all the edge switches $S_3, S_4, \dots, S_7$.}
    \label{fig:encoding2}
\end{figure}

Variable-length interface labels have to be used with some care to
ensure that the encoded paths are properly decodable. Two problems can
occur. First, the encoded paths may not be unique, i.e., two different
paths with same starting switch may be mapped into the same encoding,
thereby preventing proper routing of packets.  As an example, consider
the network and label assignment depicted on the left side of
Fig.~\ref{fig:encoding2}. In this example, we consider all possible
paths from switch $S_0$ to any of the edge switches $S_3$ through $S_7$.
Consider the path encoding $01$ at switch $S_0$.  This encoding could
result from either path $(S_0, S_4)$ or path $(S_0, S_1, S_5)$. The
switch $S_0$ has therefore no way of deciding whether to forward a
packet with encoded path $01$ to switch $S_1$ or switch $S_4$. 

A second problem is that the encoded paths may not be locally decodable.
This occurs when a switch requires global information about the entire
network in order to make the correct local forwarding decision.  As an
example, consider the same network as before, but with the label
assignment depicted on the right side of Fig.~\ref{fig:encoding2}.
Assuming the same set of paths between $S_0$ and all possible edge
switches as before, this assignment leads to unique encoded paths.
However, the encoded paths are not locally decodable. Consider for
example the path encoding $01$ at switch $S_0$. Given that $(S_0,S_2)$
is not a valid path, this encoded path is uniquely decodable to path
$(S_0,S_1,S_5)$.  Therefore, switch $S_0$ needs to forward a packet with
path $01$ to switch $S_1$. However, to make this forwarding decision,
$S_0$ needs to be aware of the collection of all possible paths in the
network. Local decodability may also be compromised if a switch needs to
know the label assignment at other switches in the network.

Both of these problems can be avoided if we impose the additional
constraint that, at every switch, the collection of labels assigned to
the interfaces of this switch form a \emph{prefix-free set}, meaning
that no label is a prefix of any other label. If this prefix-free
condition is satisfied, then each switch can make its forwarding
decision using only local information by finding the unique of its
interface labels that forms a prefix of the encoded path.  Observe that
neither of the two label assignments in Fig.~\ref{fig:encoding2} are
prefix free (since $0$ is a prefix of $01$ at switch $S_0$). On the
other hand, both the label assignment in Fig.~\ref{fig:encoding1} are
prefix free.

\subsection{Mathematical Problem Description}
\label{sec:math}

We are now ready to introduce a mathematical description of the optimal
path-encoding problem. We are given a directed graph $G = \{\mcV,
\mcA\}$ describing the network and a set of paths $\mcP$ in $G$. We are
tasked with assigning binary labels $x_a\in\{0,1\}^*$, i.e., a binary
string of arbitrary finite length, to each arc $a\in\mcA$. Denote by 
\begin{equation*}
    \ell_a \defeq \ell(x_a)
\end{equation*}
the length of the label string $x_a$. For a path $p\in\mcP$, the size
of the path encoding resulting from this assignment of labels is
\begin{equation*}
    \sum_{a\in p} \ell_a,
\end{equation*}
where the summation is over all arcs $a$ in the path $p$. 

In order to minimize the field size needed to store the path encoding,
our goal is to minimize the length of the labels for the longest (with
respect to $\ell_a$) encoded path in $\mcP$. As discussed above, we
ensure that messages are correctly routable through the network by
imposing that the set of labels $\{x_a\}_{a\in\out(v)}$ corresponding to
arcs $a$ out of any vertex $v\in\mcV$ forms a prefix-free set, meaning
that no label $x_a$ is a prefix of another label $x_{\tilde{a}}$ in the
same set.  Clearly, this prefix condition implies that the encodings of
(partial) paths with same starting vertex are unique. Moreover, it
allows each vertex $v$ to make routing decisions based on only its local
set of labels $\{x_a\}_{a\in\out(v)}$.

Consider a vertex $v$ and its outgoing arcs $\out(v)$, and consider a
set of label lengths $\{\ell_a\}_{a\in\out(v)}$. A necessary and sufficient
condition for the existence of a corresponding prefix-free set of labels
$\{x_a\}_{a\in\out(v)}$ with these lengths is that they satisfy
\emph{Kraft's inequality}
\begin{equation*}
    \sum_{a\in\out(v)}2^{-\ell_a} \leq 1, 
\end{equation*}
see, e.g., \cite[Theorem~5.2.1]{cover06}.  Moreover, for a collection of
label lengths satisfying Kraft's inequality, the corresponding set of
labels can be found efficiently as follows. Let $\Lambda$ be the largest
label length in the set. Construct a perfect binary tree of depth
$\Lambda$, with each vertex in the tree representing a binary sequence
of length up to $\Lambda$.  Next, find the arc $a$ with shortest label
length $\ell_a$. Assign this arc to the (lexicographically) first
available vertex of length $\ell_a$ in the binary tree, and remove all
its descendants from the tree. Continue this procedure with the second
shortest label length until all labels are chosen. The vertices of the
tree chosen by this procedure correspond to a set of prefix-free labels
with the specified label lengths. With this, we can focus our attention
in the following on finding the label lengths.

With this necessary and sufficient condition on the label lengths, we
can now write the optimal path-encoding problem in the following compact
form.
\begin{equation}
    \label{eq:primal_int}
    \begin{array}{lr@{\,}ll}
        \text{min} & L & & \\
        \text{s.t.} & \displaystyle\sum_{a\in p} \ell_a & \leq L, 
        & \ \forall p \in \mcP \\
        & \displaystyle\sum_{a\in\out(v)} 2^{-\ell_a} & \leq  1, 
        & \ \forall v\in \mcV \\
        & L & \in \R & \\
        & \ell_a & \in\Z, & \ \forall a\in \mcA.
    \end{array}
\end{equation}
Observe that the constraint $\sum_{a\in\out(v)}2^{-\ell_a}\leq 1$
guarantees that the optimal values of $\{\ell_a\}$ are nonnegative.
Further, since all $\ell_a$ are integers, the optimal value
of $L$ is also guaranteed to be an integer. The remainder of this paper
focuses on this minimization problem.

\begin{remark}
    \label{rem:static}
    Our assumption throughout this paper is that the path
    set $\mc{P}$ is static. In contrast, assume now that after the
    label assignment is completed and the corresponding label tables
    installed in the switches a new path $p\notin\mcP$ needs to be
    added. Even though the label assignments were not optimized for $p$,
    this new path can nonetheless be encoded using the current labels.
    Moreover, since the labels form a prefix free set at each switch,
    the resulting encoded path is uniquely routable through the network.
    Thus, enforcing the prefix-free condition has the additional
    advantage that new paths can always be added without having to
    change the label assignment. In other words, the
    network will continue to operate correctly with dynamically changing
    path set $\mc{P}$. However, since the new path $p$ was not part of
    the optimization problem yielding the label assignment, its encoded
    length may be larger than $L$. 
\end{remark}

\begin{remark}
    To alleviate the problem of newly added paths having encoded length
    larger then $L$ as mentioned in Remark~\ref{rem:static}, the set
    $\mc{P}$ should not only contain those paths that are currently
    active, but also any anticipated future paths. Such anticipated
    future paths may for example be designed to handle congestion and
    node failures.
\end{remark}

\begin{remark}
    Recall that once label lengths satisfying Kraft's inequality have
    been fixed, finding the actual labels with those lengths is
    straightforward using the algorithm described above. Hence, the
    difficult part of the optimal path-encoding problem is the
    assignment of label lengths. In the remainder of the paper, we will
    therefore focus on this subproblem of assigning label lengths with
    the understanding that the actual label assignment is then
    straightforward.
\end{remark}

\section{Main Results}
\label{sec:main}

Ideally, we would like to solve the optimal path-encoding
problem~\eqref{eq:primal_int} exactly. For special cases, such as
out-arborescences (i.e., directed trees in which the root has in-degree
zero and every other vertex has in-degree at most one), this is possible
using dynamic programming, as is illustrated in the next example.

\begin{example}
    \label{eg:tree}
    \begin{figure}[htbp]
        \centering\includegraphics{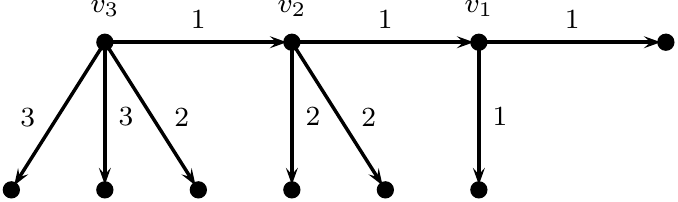}
        \caption{Graph $G$ for Example~\ref{eg:tree}. The figure shows
        next to each arc $a$ the length $\ell_a$ of the binary label $x_a$ 
        associated with that arc.} 
        \label{fig:tree}
    \end{figure}
    Consider the graph $G$ shown in Fig.~\ref{fig:tree}. The graph is an
    out-arborescence with three internal vertices labeled $v_1, v_2, v_3$ as
    shown in the figure. The set $\mcP$ consists of all paths from the
    root vertex $v_3$ to the leaves. This is the abstract
    version of the network depicted in Fig.~\ref{fig:encoding1} 
    in Section~\ref{sec:problem}.

    For out-arborescences, the optimal path-encoding problem can be
    solved exactly using dynamic programming. Consider the vertex $v_1$.
    There are two possible paths through $v_1$, one for each of its two
    children. Since these two paths have the same arcs from the root to
    $v_1$, their label lengths are the same except for the last arc.
    Since we are trying to minimize the maximum encoded path length, the
    optimal allocation of label lengths for these two arcs is $1$ for
    both of them. This in effect ``equalizes'' the two path lengths.

    Consider next vertex $v_2$. We again aim to equalize the paths
    through $v_2$. There are four such paths, one for each leaf that is
    a descendant of $v_2$. In order to equalize them, we should allocate
    a shorter length to the arc $(v_2,v_1)$ than the other two outgoing
    arcs. The optimal allocation is $1$ for the arc $(v_2,v_1)$ and $2$
    for the other two arcs. With this assignment, all four paths through
    $v_2$ have path length of $2$ from $v_2$ onward. Note that this
    assignment satisfies Kraft's inequality.

    Finally, consider the vertex $v_3$. We would again like to equalize
    paths. However, due to the integrality constraint, the best we can
    do here is to assign a label length of $1$ to the arc $(v_3,v_2)$
    and $2$, $3$, and $3$ to the other three outgoing arcs. With this,
    the maximizing path is along the topmost branch of the tree with
    length $L=3$.\footnote{For a larger example of this
    equalization, consider a vertex with six outgoing arcs of partial
    maximal encoded path lengths $5, 4, 4, 3, 2, 1$. An optimal
    label-length assignment is then $2, 3, 3, 4, 5, 6$, which satisfies
    Kraft's inequality and equalizes the partial paths to length $7$.}
    
    The example illustrates the performance improvement due to
    allocating shorter label lengths to arcs on long paths. In
    particular, if we were to assign labels of uniform length to each
    outgoing arc of a vertex, the topmost branch of the tree path would
    have encoded length $L=2+2+1=5$. This example also shows that the
    label assignment on the right in Fig.~\ref{fig:encoding1} is
    optimal.
\end{example}

While the optimal path-encoding problem can be solved exactly for
special cases as seen above, this is unfortunately not the case for
general graphs $G$ as the following theorem shows.

\begin{theorem}
    \label{thm:hardness}
    Approximating the optimal path-encoding
    problem~\eqref{eq:primal_int} by a factor less than $8/7$ is
    NP-hard.  Thus, the optimal path-encoding problem is APX-hard.
\end{theorem}

The proof of Theorem~\ref{thm:hardness} is reported in
Section~\ref{sec:proofs_hardness}. The theorem shows that at most we
should hope for an efficient constant-factor approximation algorithm
for the optimal path-encoding problem. We next describe such an
algorithm with an approximation guarantee of a factor $2$. 

Consider the relaxed version
\begin{equation}
    \label{eq:primal_relaxed}
    \begin{array}{lr@{\,}ll}
        \text{min} & L & & \\
        \text{s.t.} & \displaystyle 
        \sum_{a\in p} \ell_a & \leq L, & \ \forall p \in \mcP \\
        & \displaystyle\sum_{a\in\out(v)} 2^{-\ell_a} & \leq  1, 
        & \ \forall v\in \mcV \\
        & L & \in\R & \\
        & \ell_a & \in\R, & \ \forall a\in \mcA
    \end{array}
\end{equation}
of the integer minimization problem~\eqref{eq:primal_int}. The
function $\sum_{a\in\out(v)} 2^{-\ell_a}$ is convex in $\{\ell_a\}$, and
hence this relaxed problem is a convex minimization problem. In fact, by
rewriting Kraft's inequality as
\begin{equation*}
    \ln\biggl( \sum_{a\in\out(v)} \exp(-\ln(2)\ell_a)\biggr) \leq  0,
\end{equation*}
we see that \eqref{eq:primal_relaxed} is a \emph{geometric
program}~\cite[Section~4.5]{boyd04}. Such problems can be solved
efficiently by interior-point methods~\cite[Chapter~11]{boyd04}. 

Let $\bigl(L^{\textrm{C}}, \{\ell_a^{\textrm{C}}\}\bigr)$ be the minimizer
of the relaxed problem \eqref{eq:primal_relaxed}. Construct an integer
solution
\begin{equation*}
    \ell_a^{\textrm{I}} \defeq \ceil{\ell_a^{\textrm{C}}}, 
\end{equation*}
and set
\begin{equation}
    \label{eq:round}
    L^{\textrm{I}} \defeq \max_{p\in\mcP} \sum_{a\in p} \ell_a^{\textrm{I}}.
\end{equation}
Note that $\bigl(L^{\textrm{I}}, \{\ell_a^{\textrm{I}}\}\bigr)$ is a
valid solution of the integer path-encoding
problem~\eqref{eq:primal_int}. Moreover, the next theorem asserts that
the value $L^{\textrm{I}}$ of this solution is within a factor $2$ of
the optimal value $L^\star$ of the path-encoding
problem~\eqref{eq:primal_int}.

\begin{theorem}
    \label{thm:approximation}
    Let $L^{\textrm{I}}$ be the value of the rounded solution
    \eqref{eq:round} and let $L^\star$ be the value of the minimizer of
    the optimal path-encoding problem \eqref{eq:primal_int}. Then
    \begin{equation*}
        L^\star \leq L^{\textrm{I}} \leq 2L^\star.
    \end{equation*}
\end{theorem}

The proof of Theorem~\ref{thm:approximation} is reported in
Section~\ref{sec:proofs_approximation}. We illustrate this approximation
algorithm with a toy example.

\begin{example}
    \label{eg:integrality}
    \begin{figure}[htbp]
        \centering\includegraphics{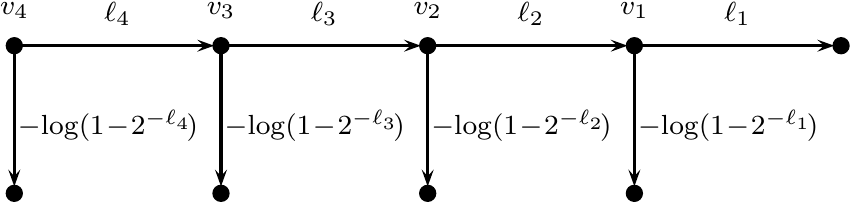} 
        \caption{Graph $G$ for Example~\ref{eg:integrality} with $K=4$.} 
        \label{fig:integrality}
    \end{figure}
    Consider the graph $G$ shown in Fig.~\ref{fig:integrality}. The
    graph is an out-arborescence consisting of $2K+1$ vertices.
    Consider the $K$ internal vertices $v_1, v_2, \dots, v_K$ forming
    the ``spine'' of the graph. Denote the label lengths of the arcs in
    this spine by $\ell_1, \ell_2, \dots, \ell_K$ as shown in
    Fig.~\ref{fig:integrality}. The set $\mcP$ is given by all the paths
    from the root to the leaves.
    
    The optimal solution for the path-encoding problem is trivial in
    this case: simply assign a value of $1$ to each arc. The resulting
    value of $L^\star$ is $K$.
    
    Let us next evaluate the relaxed problem~\eqref{eq:primal_relaxed}.
    Consider an internal vertex $v_k$ and its two outgoing arcs. Assume
    the first one has length $\ell_k$. Then the other outgoing arc has
    to have length $-\log(1-2^{-\ell_k})$ in order to satisfy Kraft's
    inequality with equality. Clearly, the optimal choice of $\ell_1$ is
    $1$. The optimal value of $\ell_2$ is given by the equation
    \begin{equation*}
        \ell_2+\log(1-2^{-\ell_2}) = -1,
    \end{equation*}
    since this equalizes the lengths of all possible paths going through
    $v_2$ as discussed in Example~\ref{eg:tree}. Using the same
    argument, we obtain the general recursion
    \begin{equation*}
        \ell_k+\log(1-2^{-\ell_k}) = -\sum_{i=1}^{k-1}\ell_i,
    \end{equation*}
    which can be solved to yield the solution
    \begin{equation*}
        \ell_k = \log(1+1/k)
    \end{equation*}
    of the relaxed problem. The resulting value of $L^\textrm{C}$ is
    \begin{equation*}
        L^\textrm{C}
        = \sum_{k=1}^K \ell_k
        = \log(K+1).
    \end{equation*}
    Comparing this to the value $L^\star = K$ of the optimal
    path-encoding problem, we see that the integrality gap is at least
    $K/\log(K+1)$, which is unbounded as $K$ increases. 
    
    Nevertheless, perhaps surprisingly, the rounded solution
    \begin{equation*}
        \ceil{\ell_k} = \ceil{\log(1+1/k)} = 1
    \end{equation*}
    of the relaxed problem is in fact equal to the optimal solution of
    the path-encoding problem for all arcs on the spine of the graph.
    Hence, $L^\textrm{I} = K = L^\star$ in this case. Thus, despite an
    unbounded integrality gap, the rounded solution is optimal in this
    case and in general yields a constant factor-$2$ approximation for
    the optimal path-encoding problem.
    
    Note that
    \begin{equation*}
        -\log(1-2^{-\ell_k}) = \log(k+1),
    \end{equation*} 
    so the rounded value $\ceil{\log(k+1)}$ on the arcs outside the
    spine of the graph is in fact considerably larger than the optimal
    solution of $1$ of the path-encoding problem. Since these arcs are not
    on the longest path, this does not affect the value of
    $L^\textrm{I}$. Nevertheless, it does indicate that, in a practical
    setting, the solution found by the rounding procedure could be
    further improved by running a local search optimization procedure.
\end{example}
 
To obtain further intuition for the solution of the relaxed
problem~\eqref{eq:primal_relaxed}, it is instructive to consider its
dual given by
\begin{equation}
    \label{eq:dual}
    \begin{array}{lr@{\,}ll}
        \text{max} & \displaystyle 
        -\sum_{a\in\mcA}\biggl(\sum_{p\ni a} \alpha_p \biggr)
        & \lefteqn{\displaystyle \log\frac{\sum_{p\ni a} \alpha_p}
        {\sum_{p\ni \tail(a)} \alpha_p}}
        & \hspace{2cm}\\[18pt]
        \text{s.t.} & \displaystyle\sum_{p\in\mc{P}} \alpha_p & = 1 &\\
        & \alpha_p & \geq 0, & \forall p\in\mcP.
    \end{array}
\end{equation}
Here, for arc $a=(v,u)$, $\tail(a)$ denotes the vertex $v$.
Moreover, we have used the shorthand notation
\begin{align*}
   \sum_{p\ni a} \alpha_p 
   & \defeq \sum_{p\in\mcP: a\in p} \alpha_p, \\
   \sum_{p\ni \tail(a)} \alpha_p
   & \defeq \sum_{p\in\mcP: \tail(a)\in p} \alpha_p,
\end{align*} 
with the nonstandard convention that a vertex $v$ is in the path $p$ if
any of its \emph{outgoing} arcs are in $p$. It is easily seen that the
relaxed primal problem~\eqref{eq:primal_relaxed} has a strictly feasible
solution, which implies that strong duality holds
\cite[Section~5.2.3]{boyd04}, i.e., the two convex programs
\eqref{eq:primal_relaxed} and \eqref{eq:dual} have the same value.  The
derivation of the dual~\eqref{eq:dual} is somewhat lengthy and can be
found in Section~\ref{sec:proofs_dual}.

Let $\{\alpha_p\}$ be a solution to the dual~\eqref{eq:dual}.
Define now a random variable $A$ taking values in $\mcA$ with 
\begin{equation*}
    \Pp(A = a) \defeq 
    \frac{\sum_{p\ni a}\alpha_p}{\sum_{p\in\mcP} \card{p}\alpha_p}
\end{equation*}
for any $a\in\mcA$, where $\card{p}$ denotes the number of arcs in the
path $p$. Furthermore, define the random variable
\begin{equation*}
    V \defeq \tail(A).
\end{equation*}
Observe that $V$ takes values in $\mcV$ with
\begin{align*}
    \Pp(V = v) 
    & = \sum_{a\in\out(v)}\Pp(A=a) \\
    & = \frac{\sum_{p\ni v}\alpha_p}{\sum_{p\in\mcP} \card{p}\alpha_p}
\end{align*}
for any $v\in\mcV$.  Finally, let $P$ be an independent random variable
taking values in $\mcP$ with
\begin{equation*}
    \Pp(P = p) \defeq \alpha_p
\end{equation*} 
for any $p\in\mcP$.

With these definitions in place, we can rewrite the dual
problem~\eqref{eq:dual} as
\begin{equation}
    \label{eq:entropy}
    \begin{array}{lr@{\,}ll}
        \text{max} & \E\bigl(\card{P(\alpha)}\bigr) 
        & \lefteqn{\displaystyle H\bigl(A(\alpha)\bigm| V(\alpha)\bigr)}
        & \hspace{2cm} \\
        \text{s.t.} & \displaystyle\sum_{p\in\mc{P}} \alpha_p & = 1 & \\
        & \alpha_p & \geq 0, & \forall p\in\mcP,
    \end{array}
\end{equation}
where we have made the dependence of the random variables on $\alpha
\defeq \{\alpha_p\}$ explicit. Here, $H(A\mid V)$ denotes the
\emph{conditional entropy} of $A$ given $V$,
\begin{align*}
    H(A\mid V)  
    & \defeq \sum_{v\in\mcV} \Pp(V=v) H(A\mid V=v) \\
    H(A\mid V = v)  
    & \defeq  \\
    & \hspace{-1cm} -\sum_{a\in\out(v)} \Pp(A=a\mid V=v) \log\Pp(A=a\mid V=v).
\end{align*}
The derivation of this entropy form of the dual problem is reported in
Section~\ref{sec:proofs_entropy}.

This reformulation of the dual has an intuitive, informal,
information-theoretic interpretation. The quantity $H(A\mid V=v)$ is
approximately (up to an additive gap of $1$) the expected length of the
optimal binary prefix-free source code for the random variable with
distribution $\{\Pp(A = a\mid V=v)\}_{a\in\mcA}$
\cite[Theorem~5.4.1]{cover06}. This distribution describes the
probability that, at vertex $v$, a path takes arc $a\in\out(v)$ under
distribution $\{\alpha_p\}$ on the paths in $\mcP$.  Averaged over all
$v$, the quantity $H(A\mid V)$ is then the average expected label
length. Since the average path contains $\E(\card{P})$ arcs, the product
$\E(\card{P}) H(A\mid V)$ can be informally understood as a
proxy for the expected size of the path encoding under this path
distribution. The dual is this quantity for the worst-case
distribution $\{\alpha_p\}$ over the paths $\mcP$. 

We can also use the dual \eqref{eq:dual} to derive a simpler projected
gradient-ascent algorithm~\cite{duchi08} for the path-encoding problem.
Recall that strong duality holds, i.e., that the two problems
\eqref{eq:primal_relaxed} and \eqref{eq:dual} have the same value.
Moreover, the derivation in Section~\ref{sec:proofs_dual} shows that the
optimal primal solution $\bigl(L^\textrm{C},
\{\ell_a^\textrm{C}\}\bigr)$ can easily be derived from the optimal dual
solution $\{\alpha_p^\textrm{C}\}$ as
\begin{subequations}
    \label{eq:primal_sol}
    \begin{align}
        \ell_a^\textrm{C} 
        & = \log\frac{\sum_{p\ni\tail(a)}\alpha_p^\textrm{C}}
        {\sum_{p\ni a}\alpha_p^\textrm{C}}, \\
        L^\textrm{C} & = \max_{p\in\mcP} \sum_{a\in p} \ell_a^\textrm{C}.
    \end{align}
\end{subequations}

The partial derivative of the dual objective function in \eqref{eq:dual}
with respect to $\alpha_p$ is proportional to
\begin{equation*}
    \Delta\alpha_p \defeq
    \sum_{a: \tail(a)\in p} 
    \frac{\sum_{\tilde{p}\ni a}\alpha_{\tilde{p}}} 
    {\sum_{\tilde{p}\ni\tail(a)}\alpha_{\tilde{p}}}
    - \sum_{a\in p}\ln
    \frac{\sum_{\tilde{p}\ni a}\alpha_{\tilde{p}}}
    {\sum_{\tilde{p}\ni\tail(a)}\alpha_{\tilde{p}}}
    - \card{p}.
\end{equation*}
This yields the following projected gradient-ascent algorithm. Start
with an initial solution 
\begin{align*}
    \alpha_p[0] & \defeq \card{\mcP}^{-1} & \forall p\in\mcP. \\
    \intertext{In iteration $t+1$ of the algorithm, set} 
    \hat{\alpha}_p[t+1] 
    & \defeq \alpha_p[t]+\gamma[t] \Delta\alpha_p[t]
    & \forall p\in\mcP, \\
    \alpha_p[t+1] 
    & \defeq 
    \bigl(\hat{\alpha}_p[t+1]-\eta[t+1]\bigr)^+
    & \forall p\in\mcP. 
\end{align*}
Here, $(x)^+ \defeq \max\{0,x\}$, and $\gamma[t]$ is a positive
parameter depending on $t$ (but not on $p$) that can be chosen using
either a line-search procedure or fixed to some small constant (see the
discussion in \cite[Section~9.3]{boyd04}). The parameter $\eta[t+1]$
needs to be chosen in each iteration such that
\begin{equation*}
    \sum_{p\in\mcP}\alpha_p[t+1] = 1,
\end{equation*}
which can be performed in $O(\card{\mcP})$ expected time as described in
\cite{duchi08}.

As was pointed out in Example~\ref{eg:integrality}, the integral primal
solution found by the rounding procedure can be further improved by
refining it with a local search optimization procedure as follows. Find
a path with longest encoding. Search along this path for any vertex at
which Kraft's inequality is not tight, and consider the arc out of this
vertex along the chosen path. Since Kraft's inequality is not tight, we
may be able to reduce the label length of this arc without violating
Kraft's inequality. If this is the case, reduce this label length by
one. Repeat these steps with different longest encoded paths until no
further reductions are possible.

Once the label lengths are found, the actual labels themselves can then
be easily derived using the algorithm described in
Section~\ref{sec:math}. We illustrate the proposed approximation
algorithm with several examples.

\begin{example}
    \label{eg:gamma}
    We applied the proposed gradient-ascent algorithm to the simple
    graph shown in Fig.~\ref{fig:tree}. With a parameter value of
    $\gamma=0.1$, the algorithm converges in $6$ steps to the optimal
    solution of the convex dual problem, from which we then recover the
    optimal solution of the convex primal problem using
    \eqref{eq:primal_sol}. The rounding of the primal solution to obtain
    a solution for the integral version needs to be done with some care,
    since numerical values (say $1.0001$, representing the value $1$)
    may be erroneously rounded up.
\end{example}

\begin{example}
    \label{eg:att}
    \begin{figure}[htbp]
        \centering \includegraphics{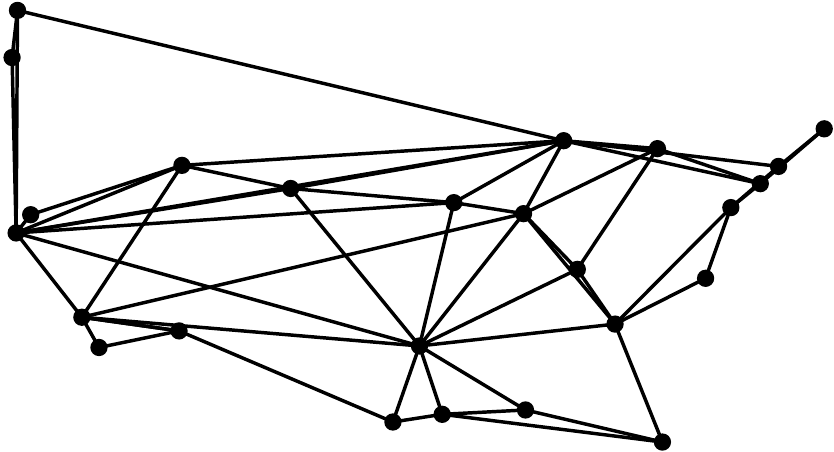}
        \caption{AT\&T MPLS backbone network from 2008.}
        \label{fig:att}
    \end{figure}
    For a more realistic scenario, we consider the AT\&T MPLS backbone
    network~\cite{att2008} as depicted in Fig.~\ref{fig:att}. This is a
    network with $25$ vertices and $224$ arcs. There are $600$ paths,
    chosen as the shortest (by hop distance) path between each ordered
    pair of vertices. A fixed-length encoding yields a maximum encoded
    path length of $15$ bits. Applying the gradient-ascent algorithm
    for variable-length path encoding proposed in this paper reduces
    this length to $10$ bits. Thus, by optimized variable-length
    encoding, the encoded path length is reduced by more than $30\%$ in
    this setting.
\end{example}

\begin{table}
    \centering
    \renewcommand{\arraystretch}{1.2}

    \begin{tabular}{@{}lrrrrr@{}}
        \toprule
        Network         & Nodes & Edges & Paths & Fixed & Variable\\
        \midrule
        3549\_3549      &  61   &  184  &  3660 &  26   &  16     \\
        4323\_4323      &  51   &  142  &  2550 &  25   &  17     \\
        Abilene         &  11   &  28   &  110  &  9    &  7      \\
        ATMnet          &  21   &  44   &  420  &  12   &  11     \\
        BBN Planet      &  27   &  56   &  702  &  14   &  10     \\
        BICS            &  33   &  96   &  1056 &  17   &  13     \\
        BT Asia Pac.    &  20   &  62   &  380  &  12   &  8      \\
        BT Europe       &  24   &  74   &  552  &  11   &  9      \\
        BT N. America   &  36   &  152  &  1260 &  16   &  12     \\
        China Telecom   &  42   &  132  &  1722 &  13   &  9      \\
        Claranet        &  15   &  36   &  210  &  9    &  7      \\
        \bottomrule
    \end{tabular}

    \caption{Comparison of maximum encoded path length under
    fixed-length interface labeling and variable-length interface
    labeling for network topologies from the RocketFuel dataset.}
    \label{tab:rocketfuel}
\end{table}

\begin{example}
    \label{eg:rocketfuel}
    We also consider several autonomous systems from the RocketFuel
    topology set~\cite{rocketfuel}. In each case, the paths are chosen 
    as in Example~\ref{eg:att}. The path lengths for both fixed and
    variable-length encodings are summarized in
    Table~\ref{tab:rocketfuel}. The average reduction in longest encoded
    path length is more than $25\%$. 
\end{example}

\section{Proofs} 
\label{sec:proofs}

\subsection{Proof of Theorem~\ref{thm:hardness}}
\label{sec:proofs_hardness}

In this section, we show that it is NP-hard to approximate the
path-encoding problem better than $8/7$ of optimal. Thus, the problem is
APX-hard. We use a reduction from {\sc (2,3)-SAT}, 
a variant of {\sc 3-SAT} that was analyzed in~\cite{tovey84} and shown
there to be NP-complete.

A Boolean expression is in conjunctive normal form (CNF) if it can be
expressed as the conjunction
\begin{equation*}
    B = C_1\wedge C_2 \wedge \dots \wedge C_M
\end{equation*} 
of clauses $C_1, C_2, \dots, C_M$. Each such clause $C_m$
is the disjunction 
\begin{equation*}
    C_m = (l_{m,1}\vee l_{m,2}\vee\dots)
\end{equation*}
of literals $l_{m,1}, l_{m,2}, \dots$. Each literal $l_{m,s}$, in turn,
is either equal to $x_n$ or its negation $\neg x_n$, where $x_1, x_2,
\dots, x_N$ are Boolean variables. In either case, we refer to $n$ as
the index $\id(l_{m,s})$ of the literal $l_{m,s}$, and we say that
variable $x_n$ is involved in clause $C_m$.

An instance of {\sc (2,3)-sat} consists of a Boolean expression $B$ in
CNF where each clause of $B$ has either $2$ or $3$ literals (with both
types of clauses present) and each variable is involved in at most 3
clauses.  Determining if an instance of {\sc (2,3)-sat} has a satisfying
assignment is NP-complete~\cite{tovey84}.  Notice that in any CNF
expression, we can assume without loss of generality that each possible
literal appears in at least one clause since otherwise the variable in
this literal can easily be removed from the expression. Thus we can
assume that in an instance of {\sc (2,3)-sat} every literal appears in
one or two clauses.

To show that the optimal path-encoding problem is APX-hard, we construct
a reduction from {\sc (2,3)-sat}. Let $I$ be an instance of
{\sc (2,3)-sat} consisting of clauses $C_1,C_2,\ldots,C_M$ over the
variables $x_1,x_2,\ldots,x_N$. From $I$ we construct an instance $L$ of
the optimal path-encoding problem consisting of a directed graph $G =
(\mcV,\mcA)$ and a set of paths $\mcP$ as follows.
 
\begin{figure}[htb]
    \centerline{\includegraphics{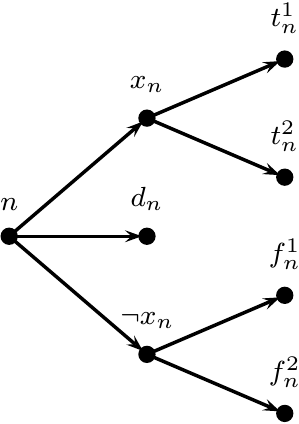}}
    \caption{Subgraph $G_n$ corresponding to variable $x_n$.}
    \label{fig:variable}
\end{figure}

We begin by defining the graph $G$. For each variable $x_n$, $1\le n\le
N$, we define the subgraphs $G_n$ of $G$ as depicted in
Fig.~\ref{fig:variable}. Subgraph $G_n$ consists of $8$ vertices labeled $n, x_n,
d_n, \neg x_n, t_n^1, t_n^2, f_n^1, f_n^2$ with arcs $(n,x_n)$,
$(n,\neg x_n)$, $(n,d_n)$, $(x_n,t_n^1)$, $(x_n,t_n^2)$, $(\neg
x_n,f_n^1)$, $(\neg x_n,f_n^2)$. The simple but crucial observation is
that at most one of the arcs $(n,x_n)$ or $(n,\neg x_n)$ in $G_n$ can be
assigned length $1$ if the lengths are to obey Kraft's inequality at
$n$.

For each variable $x_n$, $1\le n \le N$ we further define
the subgraphs $G_{N+n}$.  Subgraph $G_{N+n}$ consists of $5$ vertices
$N+n, u_{N+n}, v_{N+n}, y_{N+n}, z_{N+n}$ with arcs $(N+n,u_{N+n})$,
$(N+n,v_{N+n})$, $(u_{N+n},y_{N+n})$ and $(u_{N+n},z_{N+n})$.
Observe that the total length of the arcs $(N+n,u_{N+n})$ and 
$(u_{N+n},y_{N+n})$ is at least $2$ if
at each node the lengths of the outgoing arcs satisfy Kraft's inequality.
 
\begin{figure}[htb]
    \centerline{\includegraphics{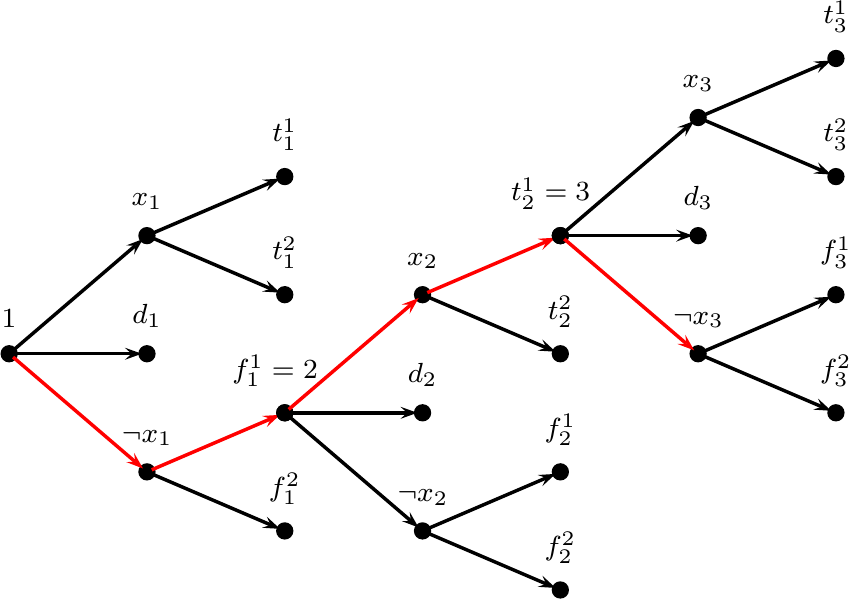}}
    \caption{Construction of graph $G$ from subgraphs $G_n$, and example
    of a path $p_m$ corresponding to clause $C_m=(\neg x_1 \vee x_2 \vee
    \neg x_3)$.}
    \label{fig:clause}
\end{figure}

We now describe how the subgraphs $G_n$ are connected to one another to
form the graph $G$. Fig.~\ref{fig:clause} illustrates the construction
for the clause $C_m=(\neg x_1 \vee x_2 \vee \neg x_3)$.

For each clause $C_m= (l_{m,1}\vee l_{m,2} \vee l_{m,3})$ containing $3$
literals, we assume without loss of generality that the variable indices
are ordered to satisfy $\id(l_{m,1})<\id(l_{m,2})<\id(l_{m,3})$.  We say
that $\id(l_{m,2})$ is a {\em successor index} of $l_{m,1}$ and that
$\id(l_{m,3})$ is a {\em successor index} of $l_{m,2}$.  For
each clause $C_m = (l_{m,1}\vee l_{m,2})$ containing $2$ literals, we
again assume without loss of generality that the variable indices are
ordered to satisfy $\id(l_{m,1})<\id(l_{m,2})$.  We say that
$\id(l_{m,2})$ is a {\em successor index} of $l_{m,1}$.  We also say
that $N+\id(l_{m,2})$ is a {\em successor index} of $l_{m,2}$.

Thus, each literal has at most two successor indices since each literal
is assumed to be in at most two clauses. We now describe
how $G$ is formed by connecting the various subgraphs $G_i$, $1\le i\le
2N$. For each literal $x_n$, $1\le n\le N$, and for each successor
index $i$ of $x_n$, we identify vertex $i$ in $G_i$  with either vertex
$t^1_n$ or vertex $t^2_n$ in $G_n$ so that if there are two successor
indices of $x_n$ then  one is identified with $t^1_n$ and the other with
$t^2_n$. Similarly for each literal $\neg x_n$ and for each successor
index $i$ of $\neg x_n$ we identify vertex $i$ in $G_i$ uniquely to one
of the vertices $f^1_n$ or $f^2_n$ in $G_n$. This
identification of vertices describes how the subgraphs $G_1,G_2,\ldots,
G_{2N}$ are connected to one another.  The example shown in
Fig.~\ref{fig:clause} illustrates how node $f^1_1$ and node $2$ are
identified as a single node and how node $t^1_2$ and node $3$ are
identified as a single node.

It remains to specify the collection of paths $\mcP$ in $G$.  Consider
clause $C_m= (l_{m,1}\vee l_{m,2} \vee l_{m,3})$ and let $i_s = \id
(l_{m,s})$, $s\in\{1,2,3\}$. For $r\in\{1,2\}$, let $p_m^r$ be the path
in $G$ from $i_r$ to $i_{r+1}$.  Then define $p_m$ as the concatenation
of the paths $p_m^1$, $p_m^2$, and arc $(i_3,l_{m,3})$. The red/gray
arcs in Fig.~\ref{fig:clause} show $p_m$ for the example clause
$C_m=(\neg x_1 \vee x_2 \vee \neg x_3)$.

Now consider a clause $C_m= (l_{m,1}\vee l_{m,2})$ 
and let $i_s = \id(l_{m,s})$, $s\in\{1,2\}$. 
Define $p_m^1$ to be the path in $G$ from $i_1$ to $i_2$.
Define $p_m^2$ to be the path in $G$ from $i_2$ to $N+i_2$.
Then let $p^3_m=(N+i_2,u_{N+i_2},y_{N+i_2})$. 
Finally, define $p_m$ to be the concatenation of $p^1_m$, $p^2_m$ and $p^3_m$. 
The set of paths is then chosen as $\mcP = \{p_1,p_2,\ldots,p_M\}$.

This completes the construction of the instance $L$ of the
optimal path-encoding problem corresponding to the instance $I$ of the
{\sc (2,3)-sat} problem.  One can easily verify that this construction
can be done in time polynomial in the size of the instance $I$.

Suppose there is a satisfying assignment $S$ for $I$.  Then for each
$x_n$ assigned the value True in $S$, give arc $(n,x_n)$ length $1$ and
arc $(n,\neg x_n)$ length $2$.  Similarly for every $x_n$ assigned the
value False in $S$ give arc $(n,\neg x_n)$ length $1$ and arc $(n,x_n)$
length $2$.  Assign length $2$ to each arc $(n,d_n)$, and assign length
$1$ to every other arc. It can easily be verified that the lengths of
the arcs out of each vertex satisfy Kraft's inequality.  For
path $p_m$ the {\em length of $p_m$}, is the sum of the lengths of the
arcs on $p_m$.  Then the length of $p_m$ is at most $7$ for all
$m\in\{1,2,\dots,M\}$ since at least one of the literals in each clause
is True, and hence the corresponding arc has length $1$. To
be more precise, for each clause $C_m$, the length of $p_m$ is $5$, $6$
or $7$ depending on whether clause $C_m$ has $3$, $2$ or $1$ true
literals respectively.  Of course, if $C_m$ only contains $2$ literals
then the length $p_m$ can only be $6$ or $7$. To summarize, if a
satisfying assignment for $I$ exists, then $L$ is at most $7$.

Conversely, suppose there is an assignment of lengths to the arcs of $G$
so that they satisfy Kraft's inequality at every vertex and such that
the length of $p_m$ is at most $7$ for $m\in\{1,2,\dots,M\}$. Then for
each $m\in\{1,2,\dots,M\}$ it is the case that if $C_m$ contains 3
literals then for at least one of the literals $l_{m,1}$, $l_{m,2}$ or
$l_{m,3}$ (or if $C_m$ contains only 2 literals then for at least one of
$l_{m,1}$ or $l_{m,2}$) the arc $(\id (l_{m,s}), l_{m,s})$ has been
assigned length 1.  Therefore the truth assignment with $x_n$ set to
False if $(n,\neg x_n)$ is assigned length $1$ and set to True otherwise
is a satisfying assignment for $I$.

Together, this argument shows that there is a solution to $L$ with
maximum path length at most $7$ if and only if there is a satisfying
assignment for $I$. Put differently, if $I$ has no satisfying assignment then
any solution to $L$ will have maximum path length at least $8$. By the
NP-hardness of {\sc (2,3)-sat}, this implies that there cannot be a
polynomial-time approximation algorithm for the optimal path-encoding
problem with approximation ratio better than $8/7$ unless
$\text{P}=\text{NP}$. \hfill\qed

\subsection{Proof of Theorem~\ref{thm:approximation}}
\label{sec:proofs_approximation}

Since
\begin{equation*}
    \sum_{a\in\out(v)} 2^{-\ell_a^{\textrm{I}}}
    \leq \sum_{a\in\out(v)} 2^{-\ell_a^{\textrm{C}}}
    \leq  1
\end{equation*}
for all $v\in \mcV$, the rounded solution $\bigl(L^{\textrm{I}},
\{\ell_a^{\textrm{I}}\}\bigr)$ is a feasible point for the integer
path-encoding problem \eqref{eq:primal_int}. The inequality $L^\star
\leq L^{\textrm{I}}$ is trivial, since $\bigl(L^{\textrm{I}},
\{\ell_a^{\textrm{I}}\}\bigr)$ is a (suboptimal) solution to 
the integer problem \eqref{eq:primal_int}. 

It remains to show that $L^{\textrm{I}} \leq 2L^\star$. Observe that the
value of a label size $\ell_a^\star$ in the optimal solution can be
equal to $0$ only if $a$ is the only outgoing arc of $\tail(a)$. But
then we can without loss of generality assume that $\ell_a^\textrm{C} =
0$ as well.  Therefore, for the path $p$ resulting in the largest path
encoding according to $\{\ell_a^{\textrm{I}}\}$, 
\begin{align*}
    L^\textrm{I}
    & = \sum_{a\in p} \ell_a^\textrm{I} \\
    & = \sum_{a\in p} \ceil{\ell_a^\textrm{C}} \\
    & \leq \sum_{a\in p} \ell_a^\textrm{C}
    + \card{ \{a\in p: \ell_a^\textrm{C} > 0 \} } \\
    & \stackrel{(a)}{\leq} L^\textrm{C} 
    + \card{ \{a\in p: \ell_a^\star > 0 \} } \\
    & \leq L^\star + \sum_{a\in p} \ell_a^\star \\
    & \leq 2L^\star,
\end{align*}
where $(a)$ follows since $L^\textrm{C}$ is the maximum of $\sum_{a\in
p} \ell_a^\textrm{C}$ over all paths, and since $\ell_a^\textrm{C} = 0$
whenever $\ell_a^\star = 0$ as argued above. This completes the
proof. \hfill\qed

\subsection{Derivation of Dual Problem~\eqref{eq:dual}}
\label{sec:proofs_dual}

We start with the Lagrangian
\begin{align*}
    & f\bigl( L, \{\ell_a\}, \{\alpha_p\}, \{\beta_v\} \bigr) \\
    & = L + \sum_{p\in\mc{P}}\alpha_p\biggl(\sum_{a\in p}\ell_a-L\biggr)
    + \sum_{v\in\mcV}\beta_v\biggl(\sum_{a\in\out(v)}2^{-\ell_a}-1\biggr) \\
    & = L\biggl(\!1\!-\!\sum_{p\in\mcP}\alpha_p\!\biggr)
    \!+\! \sum_{a\in\mcA}\ell_a\sum_{p\ni a}\alpha_p
    \!+\! \sum_{a\in\mcA}2^{-\ell_a}\beta_{\tail(a)}
    \!-\! \sum_{v\in\mcV}\beta_v.
\end{align*}
The dual is given by
\begin{equation*}
    \begin{array}{lr@{\,}ll}
        \text{max} & \displaystyle \min_{L, \{\ell_a\}} 
        f\bigl( L, & \lefteqn{ \{\ell_a\}, \{\alpha_p\}, \{\beta_v\} \bigr)}
        & \hspace{1.5cm} \\
        \text{s.t.} 
        & \alpha_p & \geq 0, & \forall p\in\mcP, \\
        & \beta_v & \geq 0, & \forall v\in\mcV.
    \end{array}
\end{equation*}

We first handle the minimization over $L$. Observe that
\begin{equation*}
    \min_{L} 
    f\bigl( L, \{\ell_a\}, \{\alpha_p\}, \{\beta_v\} \bigr)
    = -\infty
\end{equation*}
unless $1-\sum_{p\in\mcP}\alpha_p = 0$. On the other hand, if this
equality is satisfied, then the term
$L\bigl(1-\sum_{p\in\mcP}\alpha_p\bigr)$ has value $0$. Hence, the dual
can be simplified to
\begin{equation*}
    \begin{array}{lr@{\,}ll}
        \text{max} & \displaystyle \min_{\{\ell_a\}} \biggl(
        \sum_{a\in\mcA}\ell_a\sum_{p\ni a}\alpha_p
        & \lefteqn{+ \displaystyle \sum_{a\in\mcA}2^{-\ell_a}\beta_{\tail(a)}
        - \sum_{v\in\mcV}\beta_v \biggr)}  & \hspace{3.2cm} \\
        \text{s.t.} 
        & \displaystyle\sum_{p\in\mcP}\alpha_p & = 1 & \\
        & \alpha_p & \geq 0, & \forall p\in\mcP \\
        & \beta_v & \geq 0, & \forall v\in\mcV.
   \end{array}
\end{equation*}

We continue with the minimization over $\{\ell_a\}$. Taking the
derivative of the objective function with respect to $\ell_a$ and
equating to zero yields
\begin{equation*}
    \sum_{p\ni a}\alpha_p - 2^{-\ell_a}\beta_{\tail(a)}\ln 2 = 0,
\end{equation*}
which has solution
\begin{equation*}
    \ell_a = \log\frac{\beta_{\tail(a)}\ln 2}{\sum_{p\ni a}\alpha_p}.
\end{equation*}
Using this, the dual becomes
\begin{equation*}
    \begin{array}{lr@{\,}ll}
        \text{max} & \displaystyle 
        \!\!\!-\!\sum_{a\in\mcA}\biggl(\sum_{p\ni a}\alpha_p\!\biggr)
        & \lefteqn{\displaystyle \log\frac{\sum_{p\ni a}\alpha_p}
        {\beta_{\tail(a)}\ln 2 }
        \!+\! \sum_{a\in\mcA}\sum_{p\ni a}\frac{\alpha_p}{\ln 2}
        \!-\! \sum_{v\in\mcV}\beta_v}  & \hspace{4.4cm} \\
        \text{s.t.} 
        & \displaystyle\sum_{p\in\mcP}\alpha_p & = 1 & \\
        & \alpha_p & \geq 0, & \forall p\in\mcP \\
        & \beta_v & \geq 0, & \forall v\in\mcV.
   \end{array}
\end{equation*}

The maximization over the dual variables $\{\beta_v\}$ can be performed
analytically. Taking the derivative of the objective function with
respect to $\beta_v$ and equating to zero yields,
\begin{equation*}
    \sum_{a\in\out(v)}\biggl(\sum_{p\ni a}\alpha_p\biggr)
    \frac{1}{\beta_v\ln 2} - 1 = 0,
\end{equation*} 
which has solution
\begin{align*}
   \beta_v
   & = \frac{1}{\ln 2}\sum_{a\in\out(v)}\sum_{p\ni a}\alpha_p \\
   & = \frac{1}{\ln 2}\sum_{p\ni v}\alpha_p,
\end{align*}
where, as before, we use the nonstandard convention that $v\in p$ if and
only if any of its outgoing arcs are in $p$. Observe that $\alpha_p \geq
0$ for all $p\in\mcP$ implies that $\beta_v\geq 0$ as required.
Substituting the optimal value of $\{\beta_v\}$ and using that
\begin{align*}
    \sum_{a\in\mcA}\sum_{p\ni a} \alpha_p 
    & = \sum_{p\in\mcP}\alpha_p \card{\{a\in p\}} \\
    & = \sum_{p\in\mcP}\alpha_p \card{\{v\in p\}} \\
    & = \sum_{v\in\mcV}\sum_{p\ni v}\alpha_p, 
\end{align*}
the dual can be simplified to 
\begin{equation*}
    \begin{array}{lr@{\,}ll}
        \text{max} & \displaystyle 
        -\sum_{a\in\mcA}\biggl(\sum_{p\ni a} \alpha_p \biggr)
        & \lefteqn{\displaystyle \log\frac{\sum_{p\ni a} \alpha_p}
        {\sum_{p\ni \tail(a)} \alpha_p}} & \hspace{2cm} \\
        \text{s.t.} & \displaystyle\sum_{p\in\mc{P}} \alpha_p & = 1 &\\
        & \alpha_p & \geq 0, & \forall p\in\mcP,
    \end{array}
\end{equation*}
as claimed.

\subsection{Derivation of Entropy Form~\eqref{eq:entropy} of the Dual}
\label{sec:proofs_entropy}

The dual objective function can be rewritten as
\begin{align*}
    - \sum_{a\in\mcA} & \biggl(\sum_{p\ni a} \alpha_p \biggr)
    \log\frac{\sum_{p\ni a} \alpha_p}
    {\sum_{p\ni \tail(a)} \alpha_p} \\
    & = -\biggl(\sum_{p\in\mcP}\card{p}\alpha_p\biggr)
    \sum_{a\in\mcA} \Pp(A=a) \log\frac{\Pp(A=a)}{\Pp(V=\tail(a))}.
\end{align*}
Now,
\begin{align*}
    & \sum_{a\in\mcA} \Pp(A=a) \log\frac{\Pp(A=a)}{\Pp(V=\tail(a))} \\
    & = \sum_{v\in\mcV} \Pp(V=v) \!\!\!\!\!\sum_{a\in\out(v)} \!\!\!\!
    \frac{\Pp(A=a)}{\Pp(V=v)}
    \log\frac{\Pp(A=a)}{\Pp(V=v)} \\
    & = 
    \sum_{v\in\mcV} \Pp(V=v) \!\!\!\!\!\sum_{a\in\out(v)} \!\!\!\!
    \frac{\Pp(A=a,\hspace{-1pt} V=v)}{\Pp(V=v)}
    \log\frac{\Pp(A=a,\hspace{-1pt} V=v)}{\Pp(V=v)} \\
    & = 
    \sum_{v\in\mcV} \Pp(V=v) \!\!\!\!\!\sum_{a\in\out(v)} \!\!\!\!
    \Pp(A=a\!\mid\! V=v)
    \log\Pp(A=a\!\mid\! V=v) \\
    & = -H(A\mid V),
\end{align*}
so that the dual objective function becomes 
\begin{equation*}
    \E(\card{P}) H(A\mid V),
\end{equation*}
as needed to be shown.

\section{Conclusion}
\label{sec:conclusion}

We presented a mathematical formulation of the problem of minimizing
encoded paths and developed a $2$-approximation algorithm for this
problem. The algorithm allows interface labels of variable length at
each switch. Compared to a baseline fixed-length encoding, the
flexibility of this variable-length approach allows the algorithm to
yield up to a $30\%$ reduction in the length of the maximum encoded paths when
tested on real-world ISP topologies. While the problem of
path encoding was analyzed in this paper from a theoretical point of
view, in follow-up work, our proposed variable-length approach has been
implemented in the industry-standard Open vSwitch (OVS) in the Linux
kernel \cite{hari15}.

The algorithm presented in this paper assigns labels to switch
interfaces, rather than to paths. As a consequence, new paths can always
be added at any time. Moreover, these routes can be pre-optimized by
adding future projected paths as well as backup paths to the initial
list of paths when running the algorithm. An interesting open problem
for future research is how to incrementally update interface labels to
optimize the path encoding for unanticipated topology or path changes.

\end{document}